ARTICLE

# Convection and Stratification of Temperature and Concentration

## Alexey Fedyushkin*

Ishlinsky Institute for Problems in Mechanics RAS, Moscow, 119526, Russia
*Corresponding Author: Alexey Fedyushkin. Email: fai@inpmnet.ru

### ABSTRACT

The paper is devoted to the study of natural convection and the formation of delamination in an incompressible liquid due to convective laminar flows in a closed region heated from the side. Weak, medium and intensive modes of stationary laminar thermal, concentration and hemoconcentration (thermohaline) convection are considered, in which nonlinear flow features are manifested that can radically change the flow structure and characteristics of heat and mass transfer. Nonmonotonic dependences of temperature and concentration segregation in the center of the square region on the Grashof number (an intensity of flow) were found. The features of the formation of a nonstationary periodic structure of thermal convection in boundary layers and in the core of a convective flow in a closed region are shown. Details of the formation of countercurrents inside the region with the direction opposite to the main convective flow are given. The influence of vertical and horizontal vibrations on oscillatory convection is shown.

### KEYWORDS

Natural convection, stratification, segregation, numerical simulation, vibrations.

## Nomenclature

| | |
|---|---|
| g | gravitational acceleration (m/s$^2$) |
| H | height the region (m) |
| L | width of the region (m) |
| t (or time) | dimensionless time |
| Pr | Prandtl number $Pr = a / \nu$ |
| Sc | Schmidt number Sc= |
| Gr | thermal Grashof number $Gr = g\beta_T(\theta_2 - \theta_1)H^3 / \nu^2$ |
| $Gr_C$ | concentrational Grashof number $Gr_C = g\beta_C(s_2 - s_1)H^3 / D^2$ |
| Ra | Rayleigh number $Ra = Gr \cdot Pr$ |
| $Ra_c$ | concentrational Rayleigh number $Ra_c = Gr_c \cdot Sc$ |
| θ | temperature, [K] |
| s | concentration |

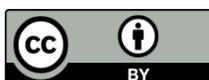

| | |
|---|---|
| T | dimensionless temperature |
| C | dimensionless concentration |
| $v_x$ | dimensionless velocity component in x direction |
| $v_y$ | dimensionless velocity component in y direction |
| $\Psi_{max}$ | maximum stream function |
| $U_{max}$ | maximum velocity $v_x$ |
| $V_{max}$ | maximum velocity $v_y$ |
| Nu | average Nusselt number on the wall |
| x, y | cartesian dimensionless coordinates |
| X, Y | cartesian coordinates (m) |
| $V_x$ | velocity component in x direction (m/s) |
| $V_y$ | velocity component in y direction (m/s) |
| a | thermal diffusivity ($m^2/s$) |
| D | diffusion coefficient ($m^2/s$) |
| $\nu$ | kinematic viscosity [$m^2/s$] |
| $\beta_T$ | thermal coefficient of volumetric expansion (1/K) |
| $\beta_c$ | concentrational coefficient of volumetric expansion |
| f | dimensionless frequency |
| A | dimensionless velocity amplitude |

**Subscripts**

| | |
|---|---|
| 1 | index of value on the left wall |
| 2 | index of value on the right wall |
| max | index of maximum value |
| T | index of thermal |
| C | index of concentrational |

## 1 Introduction

The experiments performed by Benar [1] and their theoretical interpretation by Rayleigh [2] can be considered the beginning of the study of natural convection in liquids and gases, after which almost 125 years have passed. Further research includes the works of Prandtl, Karman, Batchelor, Kutateladze, Bejan, Cormack, Imberger, Gebhard, Jaluria, Gershuni, Zhukhovitsky, Lyubimov, Tarunin, Birich, Zimin, Shaidurov, Polezhaev, Davis and their colleagues [1-50]. Unfortunately, this is an absolutely incomplete references list since there are a huge quantity of works on the study of convective processes, and this article does not have the purpose and opportunity to review them all. There are many review papers on convection heat mass transfer processes for example [44-46]. Such a lot off scientific papers is due to the variety of convective processes, fundamental interest to them, as well as the need for and importance of studying them for many applications (auto, aviation and space technology, energy (including nuclear), technologies for obtaining new materials, medicine, life support and safety systems, etc.). It should be noted that the variety of gravitational convective flows is due not only to dimensionless parameters (liquid properties, volume size and intensity of external thermal and mass fluxes), but also to the mutual direction of gravity vectors and external thermal and mass fluxes attached to volume [10]. In this article, we will consider only one case: this is a square area with horizontal fluxes of heat and mass from the vertical boundary walls. Despite the intensive study of the processes of convective heat and mass transfer, many problems remain poorly understood due to their nonlinear nature. At certain values of the determining parameters, laminar (stationary or quasi-stationary) fluid flows can exhibit nonlinear properties that can significantly change the structure of the fluid flow and the characteristics of heat and mass transfer. For example: 1) the existence of hysteresis of stationary values of the Nusselt number and the structure of the convective flow with increasing and decreasing slopes of the convective cell is known [47], 2) Effect of maximum temperature (concentration) stratification [10, 11, 25, 33, 34] and 3) well known that during vibrational action on continuous media, their anomalous nonlinear peculiarities and resonant properties may manifest themselves [16, 17, 48-50]. It must be remembered that many analytical solutions of convection problems have their own ranges of applicability. For example, based on the analysis of the equations of motion for the plane case, Batchelor [4] suggested that during convection the core is isothermal and rotates with a constant and uniform vortex of velocity, which is not always true.

In an initially homogeneous liquid located in the gravity field when heat or mass is supplied, vertical stratification in density may occur due to convective mixing. Temperature and concentration stratification in liquid volumes during convective mixing of liquids is observed in many convective processes, both in terrestrial conditions and in microgravity. Convective separation of liquid by temperature and concentration can have both positive and negative implications for various applications. The study of such heat and mass transfer processes is relevant not only from a fundamental point of view, but also for many applications. Therefore, knowledge of the patterns of formation of stationary (quasi-stationary) flow and stratification structures in liquids is important, for example, for specialists in growing single crystals in terrestrial and space conditions, since in technological processes of obtaining materials, an urgent task is to determine the possibility of regulating temperature or concentration stratification in a liquid volume in order to obtain homogeneous perfect materials with specified properties.

The occurrence of temperature or concentration stratification in the volume of liquidIn is an important aspect in convective heat and mass transfer processes [4, 5]. In addition, there is a wide range of processes and important applications where stratification with layered flows and segregation  play a decisive role, for example, in the technological processes of obtaining new perfect materials from melts and solutions [14, 15], environmental ecology [13], these are the tasks of storing and using liquid rocket fuel in tanks [10, 11]; problems of a boiling [17], problems cooling  electronic devices and safety of nuclear energy [19, 46], cleaning indoor air from pollution, smoke, as well as from finely dispersed liquid inclusions infected with viruses, in particular, COVID-19, etc [32]. Convective processes of heat and mass stratification can have large- and micro-scale character with laminar and turbulent flows and their studies are determined by different goals. It is important to consider the peculiarities of hydrodynamics and heat and mass transfer with heat and mass stratification not only in terrestrial conditions, but in microgravity conditions also [10, 12, 15, 18, 34-38]. These phenomena of stratification can be both negative and positive from the point of view of their use by a person. For example, when obtaining new materials, the macro-inhomogeneous distribution of impurities in the melt is a negative factor, since they tend to obtain single crystals with a uniform distribution of impurities across the ingot [10, 14, 15, 34-38]. Density segregation and stratification in a liquid can also have a positive aspect, for example, when separating substances or obtaining eutectic materials with a certain structure [14].

Studies of stratification and formation of stratified convective structures are carried out using experimental, analytical and numerical methods. For slow flows, there are methods with analytical solutions and approximations, for example, [7, 20-22], but experimental [23–28] and numerical methods [28-43] are needed to reproduce the flow features under intense convection.

This article demonstrates the manifestation of nonlinear features of laminar thermal and thermo-concentration convection, as well as the effect of vibration effects on the vertical stratification of temperature and impurities.

## 1. Problem Statement and Mathematical Model

The problem of gravitational thermal and concentrational convection of an incompressible liquid in a cavity with aspect ratio $L/H =1$ (were L – is length and H – is height of the calculated region), laterally heated in the field of gravity with acceleration of free fall g, is considered. At lateral heating, constant values of temperature $\theta_1$ and $\theta_2$ ($\theta_1 < \theta_2$) and concertation $s_1$ and $s_2$ on the side walls are set; for velocities, non-slip conditions are set. The following boundary conditions are considered: for velocity – the non-slip condition, for temperature on horizontal walls – line profile $T|_{y=0,\ y=1} = x$, and for concertation – no mass flow condition $\partial C/\partial y|_{y=0,\ y=1} = 0$ are set. The scheme of the calculated geometry, boundary conditions and isotherms in layer for thermal conductivity case are shown in Fig. 1.

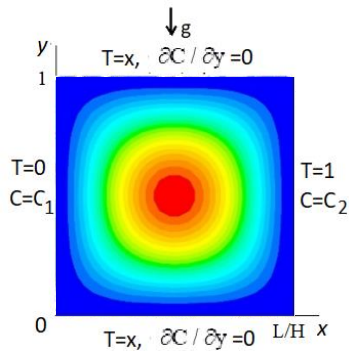

**Figure 1:** The calculated region and boundary conditions.

The mathematical model is based on the numerical solution of the unsteady 2D Navier-Stokes equations for an incompressible fluid in the Boussinesq approximation and the equations of energy and mass transfer, which in a cartesian coordinate system, in dimensionless form, in variables: $\psi$ - stream function, $\omega$ - vortex, T -temperature, C - concentration, can be written as follows [8, 10, 39]:

$$\frac{\partial^2 \psi}{\partial x^2} + \frac{\partial^2 \psi}{\partial y^2} = -\omega \tag{1}$$

$$\frac{d\omega}{dt} = \frac{\partial^2 \omega}{\partial x^2} + \frac{\partial^2 \omega}{\partial y^2} + Gr\frac{\partial T}{\partial x} + Gr_C \frac{\partial C}{\partial x} \tag{2}$$

$$\frac{dT}{dt} = \frac{1}{Pr}\left(\frac{\partial^2 T}{\partial x^2} + \frac{\partial^2 T}{\partial y^2}\right) \tag{3}$$

$$\frac{dC}{dt} = \frac{1}{Sc}\left(\frac{\partial^2 C}{\partial x^2} + \frac{\partial^2 C}{\partial y^2}\right) \tag{4}$$

where x, y – horizontal and vertical cartesian coordinates; u, v – components of the velocity vector; t – time; T – temperature; C – concentration; g – vector of gravitational acceleration of the earth's free fall directed opposite axis y; $\beta$, $\beta_C$, $\nu$, a, D – coefficients of temperature and concentration expansion of the liquid, kinematic viscosity, thermal conductivity and diffusion factor, respectively; in the future, we will use of dimensionless velocity and time (which was made dimensionless through the viscosity $\nu$ and height of the calculation region H). The problem is characterized by dimensionless parameters: the Grashof number $Gr = g\beta_T(\theta_2 - \theta_1)H^3/\nu^2$ (or Rayleigh number $Ra = Gr \cdot Pr$), concentrational Grashof number $Gr_C = g\beta_C(s_2 - s_1)H^3/D^2$ ($Ra_c = Gr_c \cdot Sc$), Prandtl number $Pr = a/\nu$, Schmidt number $Sc = \nu/D$ and aspect ratio L/H=1.

The results presented in this paper were obtained using the finite-difference scalar method [10, 39]. The good accuracy of numerical results was confirmed by comparison with experimental data and comparison of numerical results obtained by different numerical models [10, 26, 28, 40 – 43].

## 2. Benchmark of the Model on deVahl Davis Test Problem

The test problem of thermal convection of a viscous incompressible liquid (Pr=0.7) in a square closed area with thermally insulated horizontal walls and with set temperatures on vertical walls ($T_1$=1, $T_2$=0) is considered. This de Vahl Devis task was announced more than 40 years ago as an international test for computer codes. About 40 different numerical solutions to this problem have been sent by various authors. In the paper [42] "benchmark solutions" were obtained for different Rayleigh numbers by extrapolating to a zero–step grid of solutions obtained by different methods on different grids. In Table 1 Method 1 is "benchmark solution" [42, 43], method 2 is the model used in this paper with mesh 65*65 nodes. In Fig. 2 the isolines of the stream function and the isotherms of the solution of the de Vahl Devis problem for Ra=$10^3$ (left) and for R=$10^6$ (right), obtained by method 2 are shown.

**Table 1:** Comparison results of the numerical models

| Ra | Method | Nu | $\Psi_{max}$ | $U_{max}$ | $V_{max}$ |
|---|---|---|---|---|---|
| $10^3$ | 1 | 1, 118 | 1, 654 | 5, 139 | 5, 207 |
|  | 2 | 1, 119 | 1, 658 | 5, 102 | 5, 185 |
| $10^4$ | 1 | 2, 243 | 7, 142 | 22, 786 | 27, 630 |
|  | 2 | 2, 250 | 7, 167 | 22, 705 | 27, 365 |
| $10^5$ | 1 | 4, 519 | 13, 538 | 48, 915 | 96, 606 |
|  | 2 | 4, 505 | 13, 586 | 48, 850 | 97, 234 |
| $10^6$ | 1 | 8, 800 | 23, 592 | 91, 032 | 308, 958 |
|  | 2 | 8, 792 | 23, 674 | 90, 903 | 301, 777 |

Method 1 is the "benchmark solution" [42, 43], method 2 is the solution of our model using a grid of 65*65 nodes (the discrepancy is less than 3%) Fig. 2.

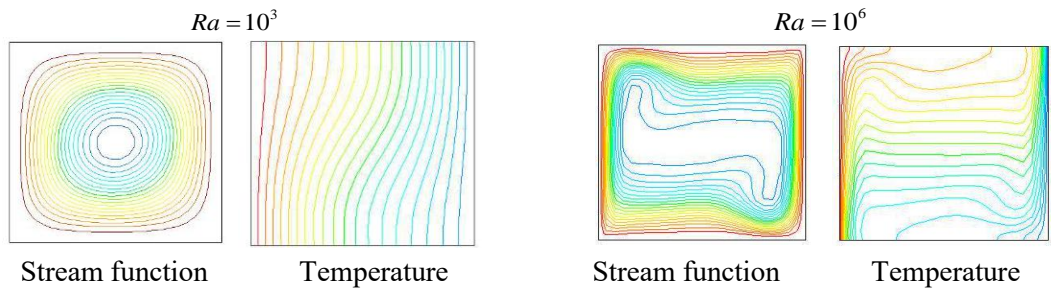

**Figure 2:** The stream function isolines and the isotherms of the solution of the de Vahl Devis problem for Ra=$10^3$ (left) and for R=$10^6$ (right)

The results simulation for large Rayleigh numbers Ra=$10^7$-$10^9$ and Pr = 5.8, for horizontal layers L/H=7-12, were compared with local experimental data on the uneven grids with 141*33 and 141*65 nodes, the comparison results showed good model accuracy and are given in [10, 28, 33].

The results of this work were obtained on an uneven grid with the number of nodes 200*200.

## 3. The Results of Numerical Simulation

Gravitational convection in a square cavity heated from the side with binary mixtures with a concentration C of a light component are considered Fig. 1. Ranges of dimensionless parameters $0 < Gr < 10^8$, $Pr = 0.7$, $0 < Gr_c < 10^8$, $10^{-2} < Sc < 10^2$ are considered: corresponding to laminar stationary and vibrational convection. The vertical stratification was estimated by the values of derivatives of temperature $(\partial T/\partial y)$ and concentration $(\partial C/\partial y)$ along the vertical y coordinate.

*3.1 Steady State Convection*

In Fig. 3 pictures of steady-state thermal convection ($Gr=10^4, 10^5, 10^6, Pr=0.7$) in the form of isolines of the stream function, isotherms and lines of equal concentration of impurity for different Schmidt numbers ($Sc=0.01, 0.1, 0.7, 10, 100$) are shown.

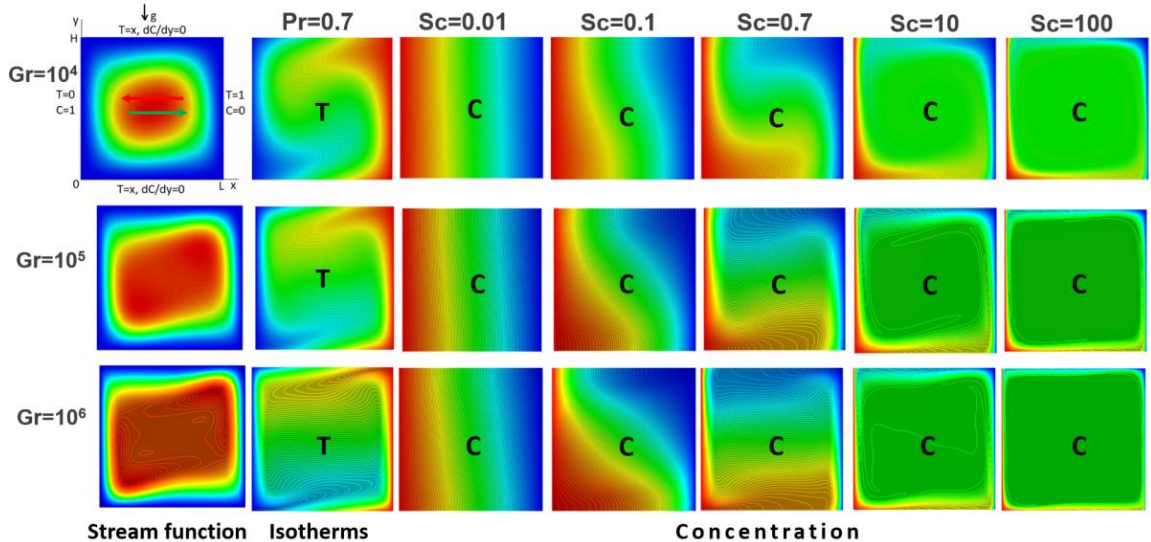

**Figure 3:** Isolines of the stream function, isotherms and lines of equal concentration of impurity for thermal convection for different Grashof ($Gr=10^4, 10^5, 10^6, Pr=0.7$) and Schmidt numbers: $Sc=0.01, 0.1, 0.7, 10, 100$.

At low Grashof numbers $0 < Gr < 10^4$ ($Pr=0.7$, $L/H=1$), the flow structure in the problem of thermal convection in a square cavity is single-vortex, with an increase in the Grashof number of more than $10^5$, secondary vortices ("cat's eyes") begin to form, which shift to the upper corner near the heated wall and to the lower near the cold one, while maintaining the diagonal symmetry of the flow.

With an increase in the Grashof number, the flow ceases to be stationary and at $Gr=10^6$ (Fig.4) the flow becomes quasi-stationary with weak periodic changes in velocity, and the secondary vortices of "cat's eyes" are formed and practically do not change (Fig. 5) [6, 7, 9]. transition on quasi-stationary mode presented in Fig. 4. In Fig. 4 on the left shows a graph with the dependencies of the average maximum and minimum temperature derivatives ($\partial T/\partial y$) along the vertical coordinate in the cross-section x=0.5 in time for $Gr=10^6$. In Fig. 4 on the right shows the isotherms and the current function in quasi-stationary mode. At $Gr=10^6$ the flow structure and temperature distribution practically do not change, although the local values of velocity and temperature undergo weak periodic oscillations [7, 9].

The formation and existence of stationary layered flow structures with countercurrents directed towards the main flow is shown in Fig. 5 for a square region ($Gr=10^6, Pr=0.7, L/H=1$). These countercurrents are formed due to intense convective flow, steady vertical stratification of density induced by convection, and the presence of vertical and horizontal walls. The presence of countercurrents during thermal convection in elongated horizontal layers, for different properties of liquids and conditions, including for semi-infinite horizontal layers, was shown in [23, 27, 28, 33].

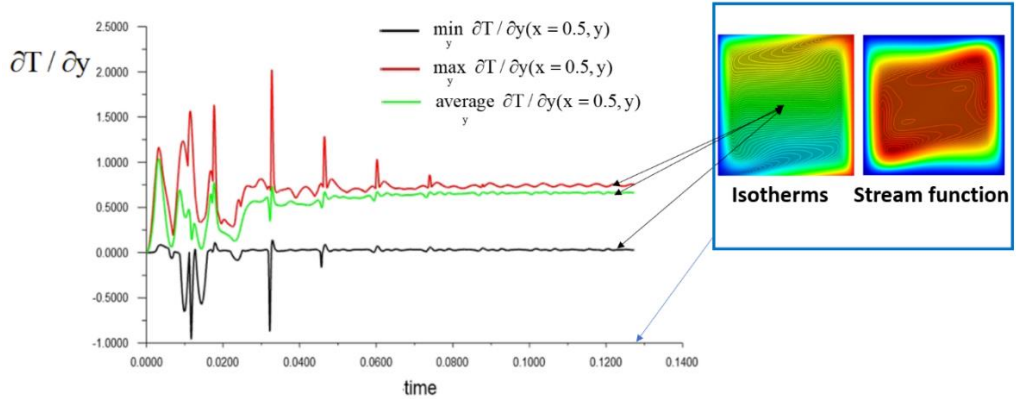

**Figure 4:** Time dependence of the derivative values $(\partial T/\partial y)$ (average, maximum and minimum values in the cross-section x=0.5) at Gr=$10^6$, Pr=0.7

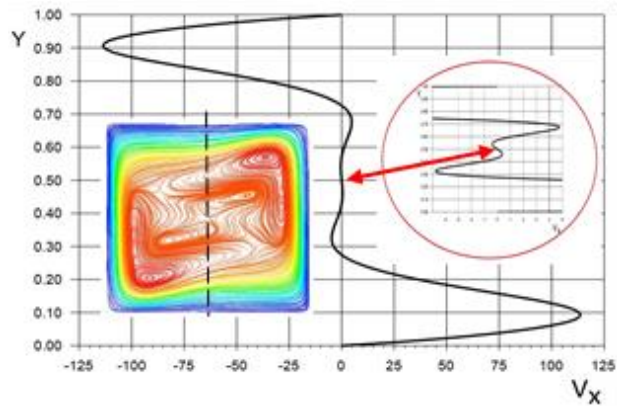

**Figure 5:** The profile of the horizontal velocity component $v_x$(x=0.5, y) in the middle vertical section for: $Gr = 10^6$, $Pr = 0.7$, $L/H = 1$. On the left are the tracks of the quasi-stationary flow; on the right is the $v_x$(x=0.5, y) profile near the center of the region on an enlarged scale.

### 3.2 Oscillatory Convection Flow

After reaching the Grashof number equal to 107, the laminar flow becomes periodically oscillatory (Fig. 6). In Fig. 6 on the left shows a graph with the dependencies of the average maximum and minimum temperature derivatives $(\partial T/\partial y)$ along the vertical coordinate in the cross-section x=0.5 in time for $Gr = 10^7$. The secondary vortices of the "cat's eyes" (which were did not move up to $Gr = 10^6$ begin to be carried away by the main convective flow (counterclockwise), these changing their intensity, splitting and uniting (Fig. 7). This manifests in the temperature field in the form of emerging thermals (thermal fingers) at the hot and cold walls (small, moving vortices appear on the walls - Tollmin–Schlichting waves, vortices increase in size as they move along vertical and horizontal walls). At $Gr = 10^7$, the entire flow pattern is periodically repeated over time. The fixed temperature on the walls contributes to the generation of vortices and the appearance of convective instability. In Fig. 7 the stream function (a - d) and isotherms (e-h) of oscillatory thermal convection for different time at a quasi-stationary mode are shown for different time moments are presented for $Gr=10^7$, $Pr=0.7$, $L/H=1$. In Fig 7a the values of the isolines of the stream function by color and the tracks during oscillatory thermal convection are shown (black line with the arrows indicate the trajectory of the moving of vortices).

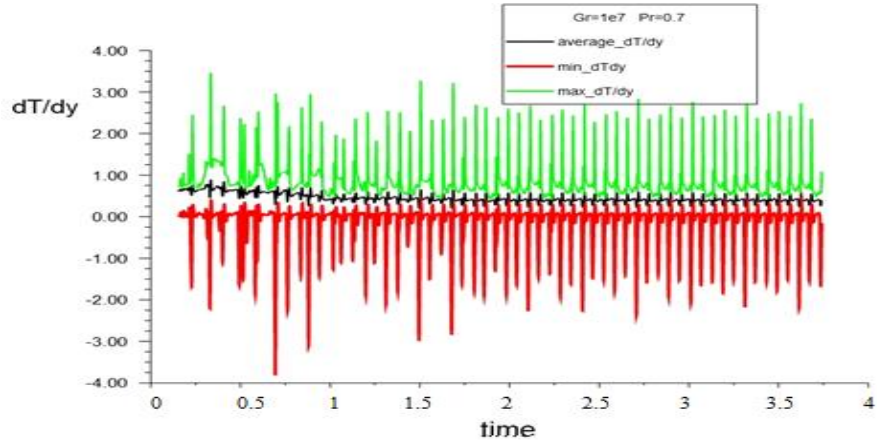

**Figure 6:** Time dependence of the derivative values $(\partial T/\partial y)$ (average, maximum and minimum values in the cross-section x=0.5) at Gr=$10^7$, Pr=0.7

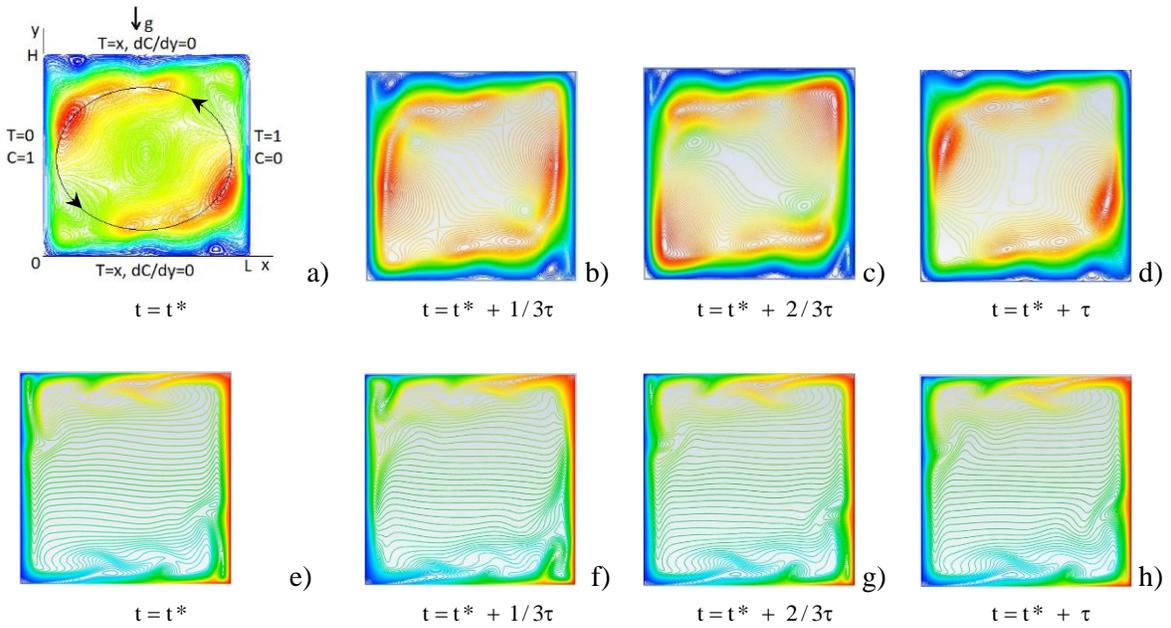

**Figure 7:** Isolines of the stream function (a - d) and isotherms (e-h) of oscillatory thermal convection for different time at a quasi-stationary mode t*=3.5 ($Gr=10^7$, $Pr=0.7$, $L/H=1$ on one oscillation period at approximately equal time intervals $1/3\tau$).

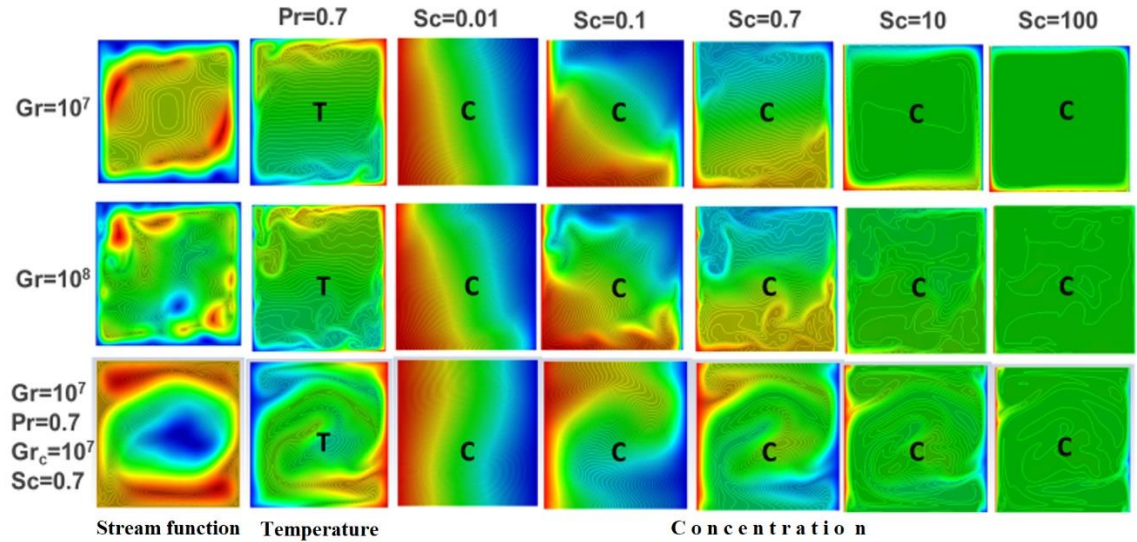

**Figure 8:** Isolines of the stream function, isotherms and lines of equal concentration of impurity for thermal convection for thermal $Gr = 10^7, 10^8$, $Pr = 0.7$ and concentrational convection $Gr=10^7, Gr_c = 10^7$, $Pr=0.7$, $Sc = 0.7$ for Schmidt numbers $Sc=0.01, 0.1, 0.7, 10, 100$.

At $Gr = 10^8$, $Pr = 0.7$, the convective flow is oscillatory, but becomes less ordered than at $Gr = 10^7$ (Fig. 8). Thermo-concentrational convection $Gr=10^7, Gr_c = 10^7$, $Pr=0.7$, $Sc = 0.7$ also has a well-defined periodic oscillatory character, but its intensity is lower than in the case of thermal convection alone ($Gr=10^7$, $Gr_c=0$) and the nature of the appearance of oscillations different than in thermal convection, The structure of thermo-concentration convection consists of two main vortices rotating in opposite directions (concentration convection causes the liquid to move clockwise; thermal convection - counterclockwise). These two main vortices are in confrontation each other, which determines the frequency of flow of this thermo-concentration convection. Fig. 8 shows that at oscillatory convection, the instantaneous concentration distributions depend on the Schmidt number and vary over time, but the average concentration fields have stationary and quasi-stationary modes.

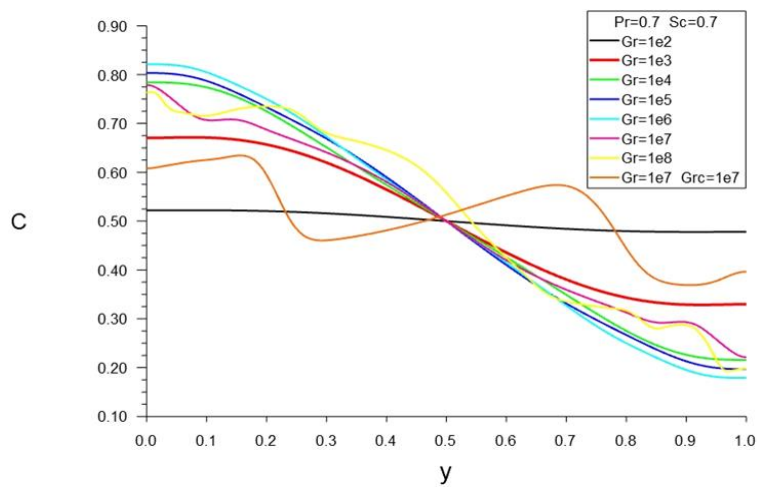

**Figure 9:** The concentration profiles in the vertical section x=0.5 for $Gr=10^2 - 10^8$ and $Gr_c = 10^7$, $Pr=0.7$, $Sc = 0.7$

The concentration profiles in the vertical section x=0.5 for thermal convection $Gr=10^2, 10^3, 10^4, 10^5, 10^6, 10^7$, Pr=0.7, Sc = 0.7 and thermo-concentrational convection $Gr=10^7$, $Gr_c=10^7$, Pr=0.7, Sc = 0.7 are shown in Fig. 9.

### *3.3 The Temperature and Concentration Stratification*

The stratification in temperature and concentration during oscillation convection varies slightly on average over time. In Fig. 10a and in Fig. 10b the dependences of the temperature derivative $\partial T / \partial y$ on the vertical coordinate calculated in the center of the square region for the Grashof number $Gr=10^3, 10^4, 10^5, 10^6, 10^7, 10^8$, Pr=0.7 for thermal and thermo-concentration convection $Gr=10^7$, $Gr_c=10^7$, Pr=0.7, Sc = 0.7 are shown. In Fig. 10 time-averaged value $\partial T / \partial y$ profiles for $Gr > 10^6$ are presented.

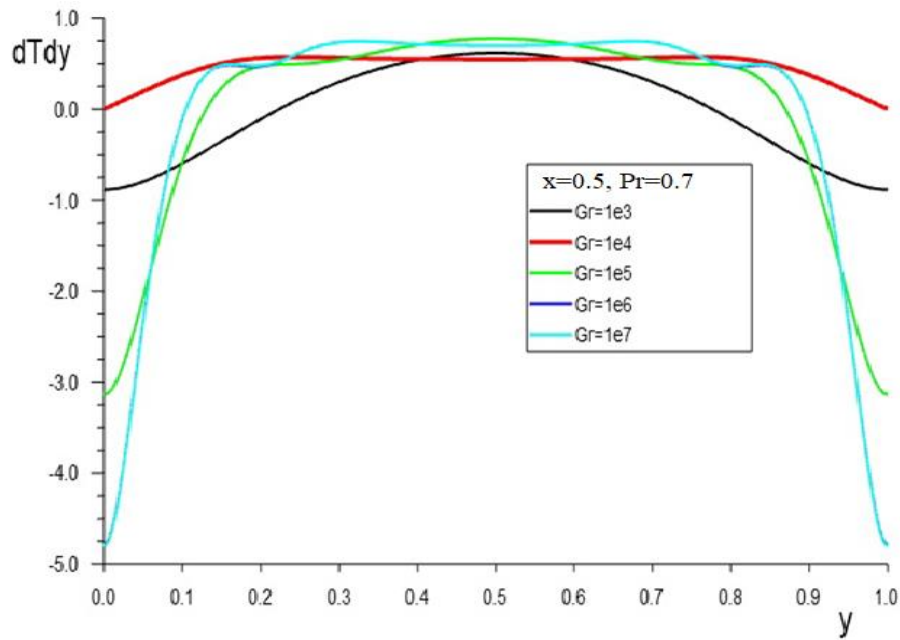

a)

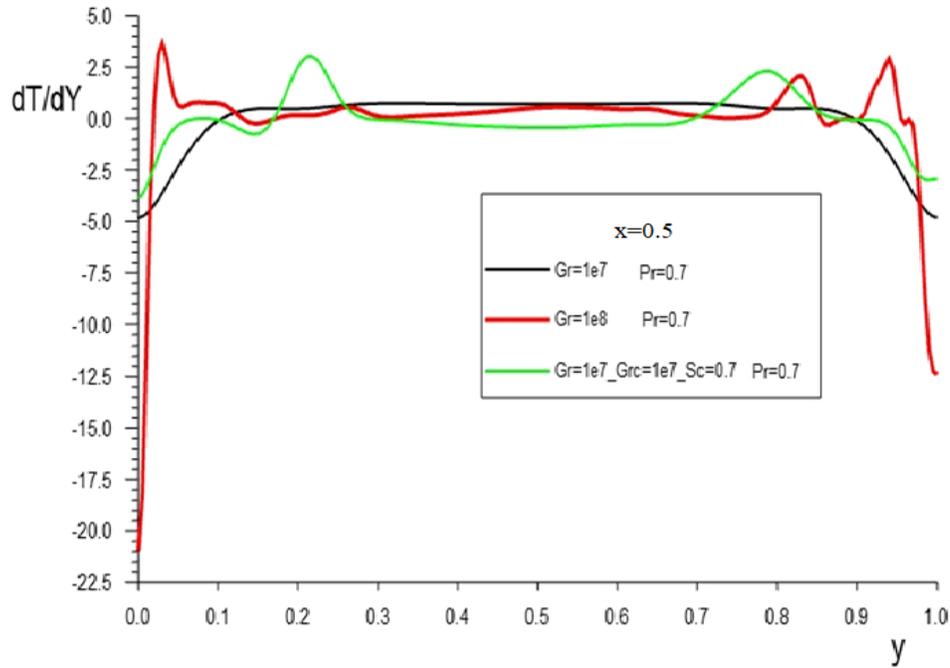

b)

**Figure 10:** The dependences of the temperature derivative $\partial T/\partial y$ on the vertical coordinate calculated in the center of the square region; a) - for the Grashof number Gr=$10^3$, $10^4$, $10^5$, $10^6$, $10^7$, $10^8$, Pr=0.7; b) - for thermal and thermo-concentration convection Gr=$10^7$, $Gr_c = 10^7$, Pr=0.7, Sc = 0.7 ($\partial T/\partial y$ was time-averaged for Gr > $10^6$).

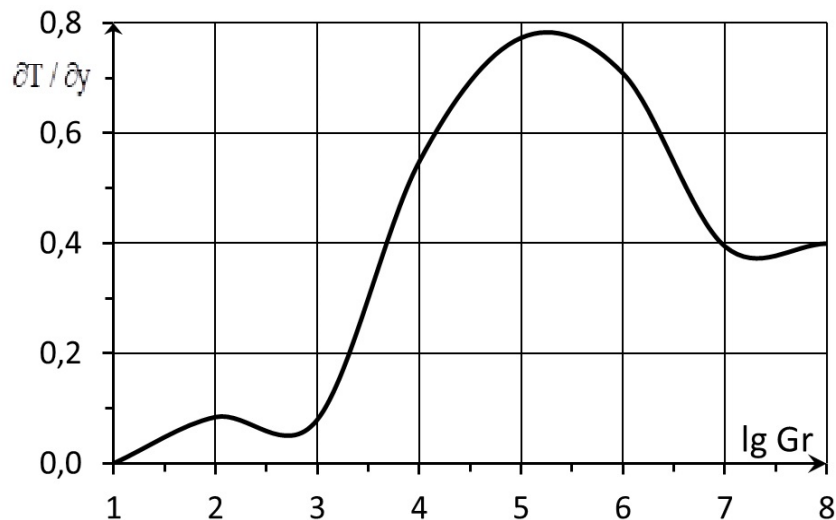

**Figure 11**: The dependence of the temperature derivative $\partial T/\partial y$ on the vertical coordinate in the center of the square area (x=0.5, y=0.5) on the Grashof number for Pr = 0.7, L/H = 1.

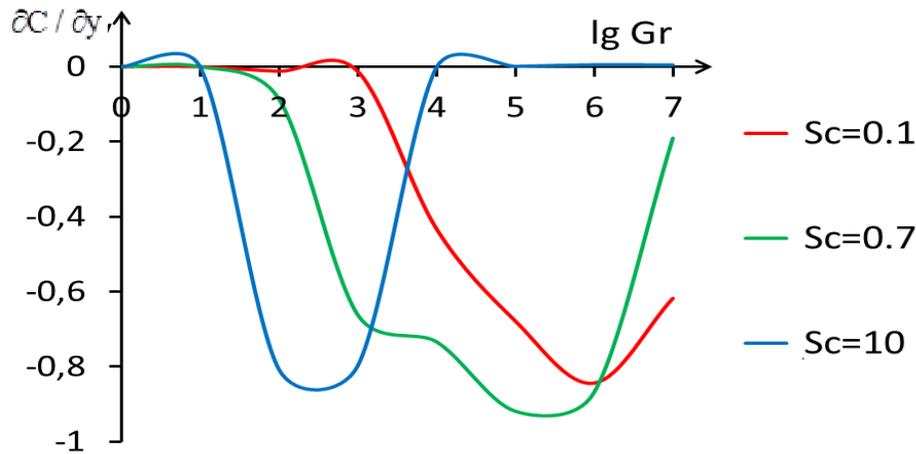

**Figure 11**: The dependence of the concentration derivative $\partial C/\partial y$ on the vertical coordinate in the center of the square area (x=0.5, y=0.5) on the Grashof number for $Pr = 0.7$; $Sc = 0.1, 0.7, 10$; $L/H = 1$.

In Fig. 11 and in Fig. 12 for thermal convection the dependences on the Grashof number of the values of the derivatives of temperature ($\partial T/\partial y$) and concentration ($\partial C/\partial y$) along the vertical coordinate calculated in the center of the region (for various Schmidt numbers: Sc=0.1, 0.7, 10) are shown ($\partial T/\partial y$ and $\partial C/\partial y$ were time-averaged for $Gr > 10^6$). The dependences on the Grashof number of vertical derivatives of temperature and concentration calculated in the center of the region (Fig. 10, 11) show that the maximum stratification of temperature and concentration exists not only between the horizontal walls - the zones of the greatest temperature and concentration change (near the boundary layers) [10, 34-38], but also in the center of the calculated region.

The dependencies of the concentration derivative $\partial C/\partial y$ on the vertical coordinate calculated in the center of the square area on the Rayleigh number for thermal, concentration and thermo-concentration convection (Pr=0.7, Sc=0.7) and comparison with experiment were presented in papers [10, 26, 33].

### 3.4 Influence Vibration on the Temperature and Concentration Stratification

Two cases of vibration effects on velocity along the normal to the walls according to the law are considered:

1) horizontal vibrations from the vertical boundaries (x=0, x=1) according to law vx=A sin(2pft),

2) vertical vibrations from the horizontal boundaries (y=0, y=1) according to law vy=A sin(2pft).

In Fig. 12 the profiles of the velocity component mean_vx averaged on time in vertical section (x=0.5) for three cases: horizontal vibrations from the vertical walls according to law vx=A sin(2pft), 2) vertical vibrations from the horizontal walls according to law vy=A sin(2pft), 3) thermal convection without vibrations are presented for A= 10, f=105, Gr=107, Pr=0.7.

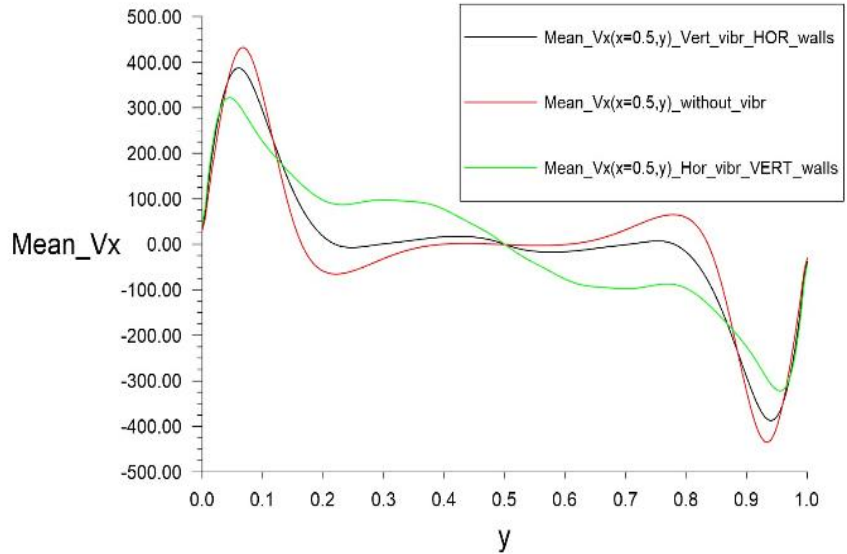

**Figure 12:** The profiles of the velocity component mean_$v_x$ averaged on time in vertical section (x=0.5) for three cases (A= 10, f=$10^5$, Gr=$10^7$, Pr=0.7):1) horizontal vibrations from the vertical walls according to law $v_x$=A sin(2πft) – green line, 2) vertical vibrations from the horizontal walls according to law $v_y$=A sin(2πft) – black line; 3) thermal convection without vibrations – red line;

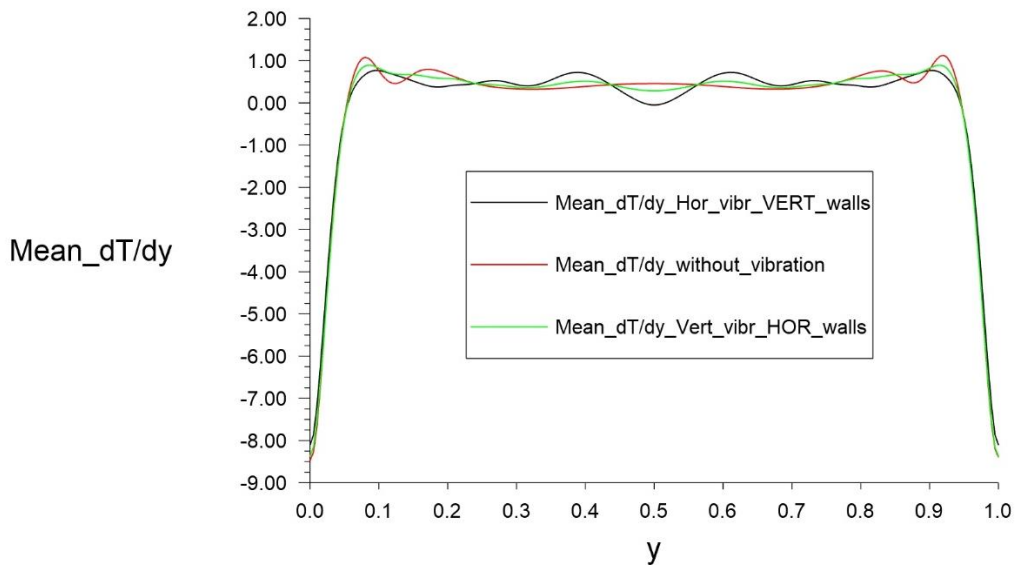

**Figure 13:** The profiles of the velocity component mean_$\partial T / \partial y$ averaged on time in vertical section (x=0.5) for three cases (A= 10, f=$10^5$, Gr=$10^7$, Pr=0.7):1) horizontal vibrations from the vertical walls according to law $v_x$=A sin(2πft) – green line, 2) vertical vibrations from the horizontal walls according to law $v_y$=A sin(2πft) – black line; 3) thermal convection without vibrations – red line;

The results shown in Fig. 12-13 show that vibrations affect the velocities both in the boundary layer and in the core of the convective cell. The convective flow averaged over time under the influence of vibrations changes its character and has a more pronounced boundary between the boundary layer and the core compared to the convective flow without vibrations.

The effect of vibrations on temperature heterogeneity is insignificant with the studied parameters of vibration exposure (A = 10, f= $10^5$) and requires further investigation.

**Conclusions**

Nonmonotonic dependences of vertical derivatives on temperature and concentration calculated in the center of the square region on the Grashof number were found, showing the presence of maximum heterogeneity of temperature and concentration depending on the Grashof number.

The pictures and difference of the formation of a nonstationary periodic structure of oscillatory thermal and thermo-concentration convection is shown.

The details of the formation of (quasi-stationary) countercurrents inside a square region directed opposite to the main convective flow are given.

The influence of vertical and horizontal vibrations on oscillatory convection is shown (Gr=$10^7$, Pr=0.7).

**Funding Statement:** This work was supported by the Russian Science Foundation grant 24-29-00101.

**Conflicts of Interest:** The author declare that he has no conflicts of interest to report regarding the present study.

**References**

1.  Benard H (1900) Revue Gen. Sei. Pures Appl. 11 1261, 1309
2.  Rayleigh 1 W S (1916) *Philos. Mag. 32* 529
3.  Prandtl L (1925) *Z. Angew. Math. Mech.*. 5 136
4.  Batchelor G. K. (1999) *An introduction to fluid dynamics*. Cambridge, U.K.; New York, NY: Cambridge University Press. 615p
5.  Gebhart B, Jaluria Y, Mahajan R. L. and Sammakia B (1988) *Buoyancy-induced flows and transport*. Hemisphere New York 1001p
6.  Gershuni G. Z., Zhukhovsitskii E. M. (1972) *Convective Stability of an Incompressible Liquid* M: GRFML 392 (In Russian).
7.  Gershuni G.Z., Zhukhovitskii, E.M. & Tarunin, E.L. (1966) Numerical investigation of convective motion in a closed cavity. *Fluid Dyn 1*, 38–42. https://doi.org/10.1007/BF01022148.
8.  Landau L D, Lifshitz E M (1987) *Course of Theoretical Physics. Fluid Mechanics 6*. Oxford: Pergamon Press 2nd ed. 556p.
9.  Tarunin E. L., Shaidurov V. G., Sharifullin A. N. (1979) Experimental and numerical study of the stability of a closed convective boundary layer. *Convective flows and hydrodynamic stability. Collection of articles. Sverdlovsk, (UNC A.N. USSR),* pp. 3-16. (In Russian).
10. Polezhaev V I, Bello M S, Verezub N A et al (1991) *Convective Processes in Weightlessness* M: «Nauka» 240 (in Russian).
11. Polezhaev V. I (1974) Effect of maximum temperature stratification and its *applications Dokl. USSR Academy of Sciences 218*, 783–786 (In Russian).
12. Fedyushkin A I (2020) Effect of Convection on Crystal Growth of Calcium Phosphate in a Thermostat under Terrestrial and Space Conditions *Fluid Dynamics 55 issue 4* 35-46 DOI: 10.1134/s0015462820040047
13. Kirdyashkin A. G. (1989) *Thermal gravitational flows and heat transfer in the asthenosphere*.

Novosibirsk: "Nauka" Siberian branch, 81p (In Russian).
14. Laudise R. and Parker R. (1974) *The Growth of Single Crystals; Crystal Growth Mechanisms: Energetics, Kinematics and Transport* Prentice-Hall New York (MIR Moscow) 540p. (In Russian).
15. Muller G. (1989) *Convection and inhomogeneities in crystal growth from the melt* Crystals 12 *Growth, Propeties, and Applications* Springer-Verlage Berlin 1–133.
16. Fedyushkin A.I. (2021) The effect of controlled vibrations on Rayleigh-Benard convection. J. of Physics: Conf. Ser. (2057)1, 012012. DOI: 10.1088/1742-6596/2057/1/012012.
17. Fedyushkin A. I. (2020) *Numerical simulation of gas-liquid flows and boiling under effect of vibrations and gravity* J. Phys.: Conf. Ser. 1479 012094 Applied Mathematics Computational Science and Mechanics: Current Problems DOI:10.1088/1742-6596/1479/1/012094.
18. Fedyushkin A. I., Burago N. G., Puntus A. A. (2019) Effect of rotation on impurity distribution in crystal growth by Bridgman method. *Journal of Physics: Conference Series*. 1359 012045. DOI:10.1088/1742-6596/1359/1/0120451.
19. Grigorchuk D. G., Strizhev V. F., Filippov A. S. (2008) *A numerical investigation of heat transfer of stratified melt with volumetric heat release in the bottom layer* Heat and Mass Transfer and Physical Gasdynamics. 46 3 386–392.
20. Gershuni G Z, Zhukhovitsky E M (1970) *On slow flow of viscous fluids in a closed region* In: «Hydrodynamics» issue 2 Perm, pp. 207-217 (In Russian).
21. Birikh R. V. Thermocapillary convection in a horizontal layer of liquid. *J. Appl. Mech. Tech. Phys. 1966 7 3* 43–44.
22. Cormack D E, Leal L G, Imberger J (1974) *J. Fluid Mech 65* 209–229.
23. Zimin V D, Lyakhov Yu N, Shaidurov G F (1971) *Experimental study of the temperature field under natural convection of a liquid in a closed rectangular cavity* In: «Hydrodynamics» 248 3 Perm pp.126-138. (In Russian).
24. Cormack D E, Leal L G, Seinfield J H (1974) *J. Fluid Mech. 65,* 231–246.
25. Bejan A, Al-Homoud A A, Imberger J (1981) *J. Fluid Mech. 109* 283–299.
26. Pshenichnikov A F, Piniagin A Yu, Polezhaev V I, Fedyushkin A I, Shaidurov G F (1985). *Thermoconcentration convection in a rectangular area with lateral heat and mass flows* Preprint. UNC of the USSR Academy of Sciences Sverdlovsk. 53p (in Russian).
27. Kirdyashkin A.G. (1984) Thermogravitational and thermocapillary flows in a horizontal liquid layer under the conditions of a horizontal temperature gradient. *Int. J. Heat Mass Transfer*. Vol. 27, N8.
28. Kirdyashkin A G, Polezhaev V I and Fedyushkin A I (1983) Thermal convection in a horizontal layer with lateral heating. *Journal of Appl. Mech. Tech. Phys. 24* 876–882.
29. Drummond J E, Korpella S A (1987) *J. Fluid Mech. 182,* 543-564.
30. Fedyushkin A I, Puntus A A (2018) Nonlinear features of laminar fluid flows on Earth and in zero gravity. *Trudy MAI 102* 20 (In Russian).
31. Fedyushkin A. I. (2020) Natural convection in horizontal layers with lateral heating at various Rayleigh and Prandtl numbers. *IOP Conference Series: Materials Science and Engineering. Vol. 927.* 012054. DOI: 10.1088/1757-899x/927/1/012054
32. Rozhkov A., Prunet-Foch B., Fedyushkin A., Vignes-Adler M. (2023) Fragmentation of water drops in collision with a small obstacle. *Atomization and Sprays. V. 33. No 10,* pp. 1-15.
33. Fedyushkin A. I. (2023) *Stratification and segregation under laminar convection.* Advanced Hydrodynamics Problems in Earth Sciences (eds. Chaplina T.), pp. 153–169, Springer: Switzerland. https://doi.org/10.1007/978-3-031-23050-9_14.
34. Polezhaev V I, Fedyushkin A I (1980) Hydrodynamic effects of concentration stratification in closed spaces. *Journal Fluid Dynamics 15 3* 331-337.
35. Nikitin S. A., Polezhaev V. I., Fedyushkin A. I. (1981) Mathematical simulation of impurity distribution in crystals prepared under microgravity conditions *Journal of Crystal Growth 52 Part 1* 471-477.
36. Nikitin S. A., Polezhaev V. I. and Fedyushkin A. I. (1981) Mathematical simulation of impurity distribution in space processing experiments with semiconductors *Advances in Space Research 1 Issue 5* 37-40.
37. Polezhaev V. I., Dubovik K. G., Nikitin S. A., Prostomolotov A. I. and Fedyushkin A. I. (1981) Convection during crystal growth on earth and in space *Journal of Crystal Growth 52 Part 1* 465-470.
38. Brown R A (1988) Theory of transport processes in single crystal growth from the melt, *AIChE Journal 34* 881.

39. Polezhaev V. I., Bune A. V., Verezub N A, et al. (1987) *Mathematical modeling of convective heat and mass transfer based on Navier-Stokes equations* M.: «Nauka» 272p (in Russian).
40. Fedyushkin A. I. (1990) *Research of a matrix method for solving convection equations. Complex of programs "MARENA"* Preprint 471 M.: IPM of the USSR Academy of Sciences 32p. (in Russian).
41. Patankar S. V. (1980) *Numerical Heat Transfer and Fluid Flow* Hemisphere 214p.
42. Davis G de Vahl and Jones T. P. (1983) Natural convection in a square cavity: a comparison exercise *Int. journal for numerical methods in fluids*, *vol. 3*, 227-248.
43. Davis G de Vahl (1983) Natural convection of air in a square cavity a benchmark numerical solution *Int. journal for numerical methods in fluids, vol.3*, 249-264.
44. Andreev, V.K., Bekezhanova, V.B. (2013) Stability of non-isothermal fluids (Review). J Appl Mech Tech Phy 54, 171–184. https://doi.org/10.1134/S0021894413020016.
45. Getling A. V. (1991) Formation of spatial structures in Rayleigh–Bénard convection. *Soviet Physics Uspekhi, Volume 34, Number 9, Sov. Phys. Usp. 34* 737. DOI 10.1070/PU1991v034n09ABEH002470.
46. L A Bol'shov, P S Kondratenko, V F Strizhov. (2001) Natural convection in heat-generating fluids. Physics Uspekhi 44 (10) 999 – 1016. DOI: 10.1070/PU2001v044n10ABEH001012
47. Pivovarov D. E. (2013) Three-dimensional convective interactions in an inclined longitudinal air layer. *Fluid Dynamics. Vol. 48, no. 3*. 321–329. DOI: 10.1134/S0015462813030058
48. Chelomei V.N. Mechanical paradoxes caused by vibrations. *Dokl. Akad. Nauk SSSR, 1983, V. 270, No.1*, pp.62-67 (In Russian).
49. Blekhman I.I., Blechman L.I., Weisberg L.A. [et al.]"Abnormal" phenomena in a liquid under the action of vibration. *Dokl. Akad. Nauk SSSR  2008. Vol. 422, No. 4,* pp. 470-474. (In Russian).
50. Ganiev R.F., Ukrainsky Ya.E. Dynamics of particles under the influence of vibration. *Kiev: Naukova dumka, 1975,* 168 p. (In Russian).